# Study of the Effects of High-Energy Proton Beams on Escherichia Coli

**Jeong Chan Park\*, Myung-Hwan Jung**

*Korea Multi-purpose Accelerator Complex, Korea Atomic Energy Research Institute, Gyeongju, Korea.*

Antibiotic-resistant bacterial infection becomes one of the most serious risks to public health care today. However, discouragingly, the development of new antibiotics has been little progressed over the last decade. There is an urgent need of the alternative approaches to treat the antibiotic-resistant bacteria. The novel methods, which include photothermal therapy based on gold nano-materials and ionizing radiation such as X-rays and gamma rays, have been reported. Studies of the effects of high-energy proton radiation on bacteria are mainly focused on Bacillus species and its spores. The effect of proton beams on Escherichia coli (E. coli) has been limitedly reported. The Escherichia coli is an important biological tool to obtain the metabolic and genetic information and also a common model microorganism for studying toxicity and antimicrobial activity. In addition, E. coli is a common bacterium in the intestinal tract of mammals.

Herein, the morphological and physiological changes of E. coli after proton irradiation were investigated. The diluted solutions of the cells were used for proton beam radiation. LB agar plates were used to count the number of colonies formed. The growing profile of the cells was monitored by optical density at 600 nm. The morphology of the irradiated cells was analyzed with optical microscope. Microarray analysis was performed to examine the gene expression changes between irradiated samples and control samples without irradiation.

E-mail: goodnews1979@gmail.com

Fax:+82-54-750-5304

# I. INTRODUCTION

Treating antibiotic-resistant bacterial infection becomes issued worldwide. It has been evaluated that around sixty percent of nosocomial infections in the United States have been caused by antibiotic-resistant bacteria [1]. However, discouragingly, the development of new antibiotics has been little progressed over the last decade. The novel methods including photothermal therapy based on gold nanoeggs [2] and gold nanorods [3] have been reported to treat antibiotic-resistant bacteria. Another approaches using ionizing radiation such as X-rays irradiation combined with nanomaterials [4] and gamma rays [5], have been employed to kill bacterial cells.

Escherichia coli (E. coli) is an important biological model to obtain metabolic and genetic information for studying toxicity and antimicrobial activity of pathogenic bacteria [6]. Most studies on the effect of ionizing radiation on E. coli have done with gamma irradiation, UV light and X-ray radiation. E. coli O157:H7, which is a food-borne pathogen, was studied on the effect of ionizing radiation with $^{137}$Cs as a gamma irradiation source [5]. Survival curve of E. coli to UV irradiation was studied during log and stationary phases of the growth [7]. It has been demonstrated that DNA degradation in E. coli induced by proton beams with different energies ranges from 1.3 to 4.75 MeV [8].

However, no systematic study has been reported on the effect of high-energy proton beams on E. coli. Herein, E. coli was exposed to spread out Bragg peak (SOBP) protons (45 MeV) at doses of 13, 23, 46 and 93 Gy.

# II. EXPERIMENTS

**Bacteria culture**

E. coli DH5α, which was purchased from real-biotech, was cultured overnight on Luria Bertani (LB) medium at 36℃ with agitation. The overnight cultured LB medium containing

E. coli was suspended in 10 mL fresh LB solution contained in 50 mL capacity conical flask. The initial optical density of E. coli cells was ~0.2 before proton irradiation.

**Proton beam irradiation**

The proton beam energy was set to 45 MeV. A spread-out Bragg peak (SOBP) dose distribution was created by passing a ridge filter type modulator designed to obtain a uniform dose distribution [9]. Eppendorf tubes containing E. coli culture medium with an initial $OD_{600}$ of around 0.2 were irradiated with dose ranging from 13 to 93 Gy. Control samples without proton irradiation were maintained under the same condition as experimental samples except irradiation.

**Growth monitoring**

Without proton irradiation, bacterial cells were monitored for 57 h by measuring the culture turbidity spectrophotometrically to obtain $OD_{600}$ values, optical density at 600 nm. After irradiation, 0.1 mL of the radiation-exposed bacteria solution were reseeded in 5 mL of fresh LB medium and allowed to grow to monitor the growth with Shimadzu UV-2550 spectrophotometer. The proton beam exposed cell suspensions and control samples were serially diluted with the ratio of 1:10,000 and 1:1,000. Aliquots of 0.2 mL bacterial solution were plated on LB agar plates to count the colonies.

**Microscopic and microarray analysis**

Control samples without proton irradiation and irradiated bacteria were taken to examine with optical microscope. Before taking, the samples were shaken up to make a uniform suspension. 200 uL of the solutions were dropped on the microscope slides and then covered with cover slip. Microscopic analysis was performed with DM2500 (Leica) at a magnification of 400× to observe the morphology change of E. coli cells. To examine the gene expression changes between irradiated samples and control samples without irradiation, microarray analysis was performed by Macrogen Inc.

### III. RESULTS AND DISCUSSION

To monitor the growth rate of E. coli under our experimental conditions, 0.1 mL of the overnight cultured E. coli solution was inoculated in 5 mL of fresh LB medium and measured the optical density at 600 nm for 57 h. It was observed that the bacterial cells reach in stationary phase after 24 h inoculation (Fig. 1A). After proton irradiation at 13, 23, 46 and 93 Gy, 0.1 mL of the irradiated bacteria solution was inoculated in 5 mL of fresh LB culture medium and measured the turbidity over 24 h with UV spectrophotometer at 600 nm. As shown in Fig. 1B, no significant differences observed at 20 min for all dose ranges. The populations of the bacterial cells with protons at 13 and 23 Gy began to increase slowly after 1 h and 1 h 40 min incubation. In case of control samples without irradiation, there was a steady increase in the populations after 1 h and 1 h 40 min incubation. On the other hand, until 1 h 40 min incubation time there was no increase in the populations at 46 and 93 Gy. As shown in Fig. 1C, when the turbidity was measured until 24 h, $OD_{600}$ values reached approximately 1.7 at all dose ranges, 13, 23, 46 and 93 Gy including control samples without irradiation. The irradiated bacteria were still able to proliferate and the growth curve had the same appearance compared with the cells without irradiation. It might be that irradiation caused the death of some bacteria but not all, and the viable bacteria proliferated as non-irradiated bacteria.

---------------------------------------------------------------------

Figure 1

---------------------------------------------------------------------

To check the ability of the colony forming of the cells, the bacterial cell solutions irradiated with protons at 0, 13, 23, 46 and 93 Gy were serially diluted with the ratio of 1:10,000 and 1:1,000, respectively. Aliquots of 0.2 mL the bacteria medium were plated on LB agar plates. After 24 h incubation, the colony formed units (CFU) in each plate were evaluated. The survival rate was calculated through $N/N_0$, where N indicates the CFU after each irradiation ranging 13, 23, 46 and 93 Gy and $N_0$ represents the CFU of the samples not exposed to irradiation. As shown in Fig.2, the survival fraction was decreased to less than 0.6 even at dose 13 Gy. E. coli O157:H7, which is a food-borne pathogen, was exposed to gamma irradiation with dose ranging from 0.5 to 3 kGy [5]. In that literature, no survivors (less than 10 CFU per grams) were found in samples exposed to the high dose of gamma irradiation. Colony-forming ability of E. coli exposed to 2.75 MeV protons was examined [8]. The 37% survival dose in that literature was $2.2 \times 10^9$ protons/cm$^2$. In our experiment conditions, The 56% survival dose was 13 Gy corresponding to $4.8 \times 10^9$ protons/cm$^2$. The 26% survival dose was 46 Gy corresponding to $8.5 \times 10^9$ protons/cm$^2$. DNA damage in E. coli and mammalian cells is dependent on Linear Energy Transfer (LET) of protons. Though the mechanism is still unclear, most probably direct or indirect damages in DNA strand are caused by proton beams and reactive oxygen species generated from the ionization of water molecules. The comparison between turbidity measurement and CFU counting showed that CFU counting was more efficient ways to recognize the change of populations of bacterial cells at initial stage.

---------------------------------------------------------------------

Figure 2

---------------------------------------------------------------------

As shown in Fig. 3A, the optical microscopic images showed that E. coli non-exposed to proton beams was visible with shape of a short rod. It was observed that E. coli cells elongated after proton irradiation with dose ranging from 13 to 93 Gy (Fig. 3B-D). The E. coli cells also elongated when exposed to UV lights and low dose of protons (0.4 and 0.7 Gy). On the other hand, the survival curve of the E. coli exposed to doses ranging from 1 to 5 Gy of protons showed no significant decrease [10]. While the survival fractions were approximately 0.57, 0.26 and 0.004 at dose of 13, 46 and 93 Gy, respectively. Usually the elongation phenomenon of E. coli has been reported when E. coli was exposed to low doses (1-50 cGy) of radiation [11]. The elongation may be resulted from working of the protective system of the cells exposed to the putative stimulus, which so-called hormesis phenomenon.

---------------------------------------------------------------

Figure 3

---------------------------------------------------------------

To check the gene expression changes between irradiated samples and control samples without irradiation, messenger RNA (mRNA) was analyzed using the Affymetrix GeneChip by Macrogen Inc. Changes in gene expression of all the probe sets on the Affymetrix GeneChip are plotted as shown in Fig. 4.

---------------------------------------------------------------

Figure 4

---------------------------------------------------------------

The up-regulated genes with greater than 2.0 fold changes in the irradiated sample (93 Gy) compared with control sample were identified (Table 1). The number of genes with increased signals greater than two fold was 22.

---------------------------------------------------------------

Table 1

----------------------------------------------------------------

We employed E. coli bacteria for proton treatment studies. E. coli is a common model organism among various microorganisms due to its safety and easy accessibility. The survival rate of microorganisms with gamma radiation has been intensively studied for food irradiation. However, it has been limitedly reported that the effect of MeV protons on bacteria. 3.2 MeV proton irradiation was performed with bacteria in atmosphere conditions for mutation studies [12]. Although several keV heavy ions irradiations are efficient for mutation because of their high linear energy transfer (LET), the experiment should be done under vacuum environments, which cause severe damages to the experimental bacteria. SOBP protons (45 MeV) can be applied to E. coli to generate mutation [13]. Bacillus subtilis spores, which have shown the resistance to extreme environment, exposed to 218 MeV protons to study the effect of high energetic proton irradiation on the spore survival [14]. High-energy proton beams may be applied to animal models bearing drug-resistant bacteria for radiation therapy.

## IV. CONCLUSION

E. coli, which is a common model organism among various microorganisms, exposed to high-energy proton beams at dose of 13, 23, 46 and 93 Gy, respectively. After proton irradiation, turbidity measurement with UV spectrophotometer and CFU analysis were performed to monitor the growth and check the populations of the bacterial cells. It was able to observe that the morphology change of E. coli cells exposed at even lethal dose of proton irradiation. 22 genes were up-regulated, more than twofold in proton irradiated samples (93 Gy) compared with unexposed samples.

**Acknowledgment**

This work has been supported in part by KOMAC (KOrea of Multi-purpose Accelerator Complex) and the Creative Research Program of KAERI.

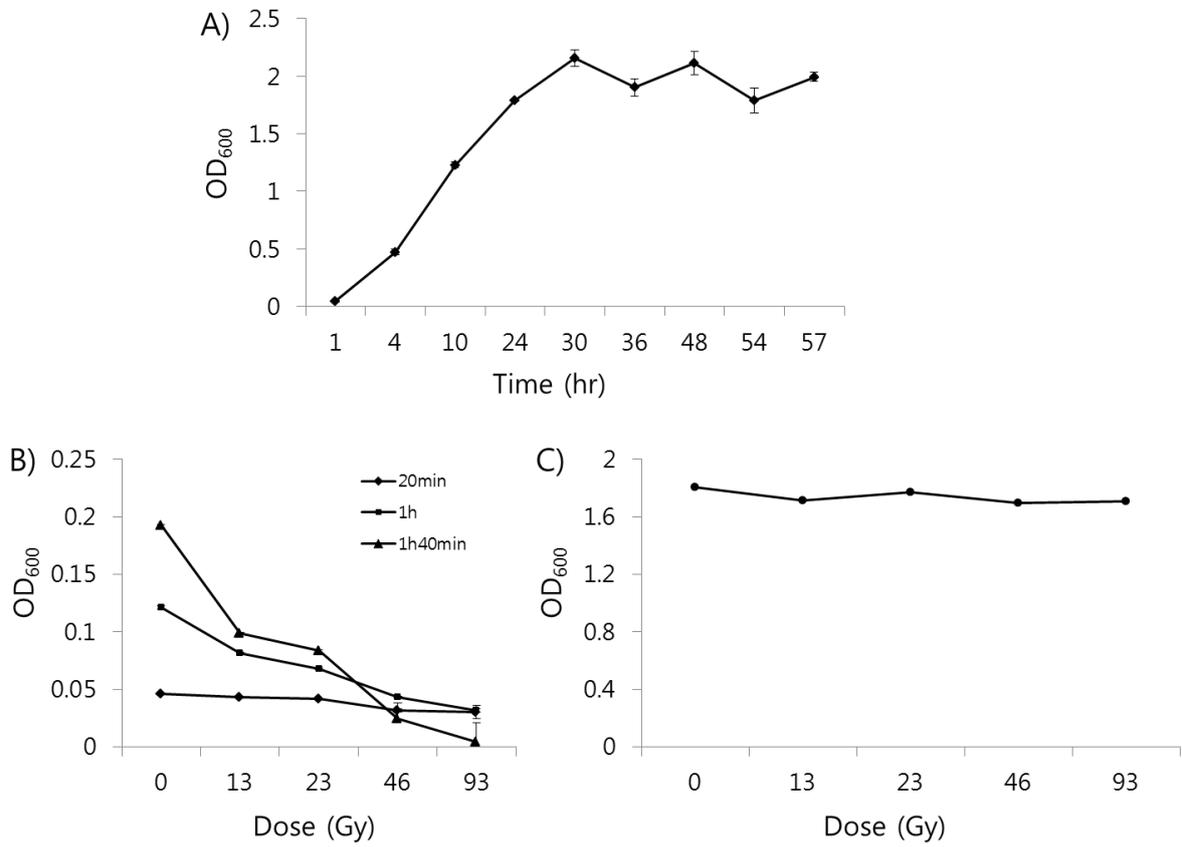

Figure 1. (A) Growth curve of E. coli for 57 h without proton irradiation. (B) Growth curve of E. coli exposed to proton beams at dose from 0 to 93 Gy. (C) 24 h-growth monitoring of E. coli submitted to proton beams at dose from 0 to 93 Gy.

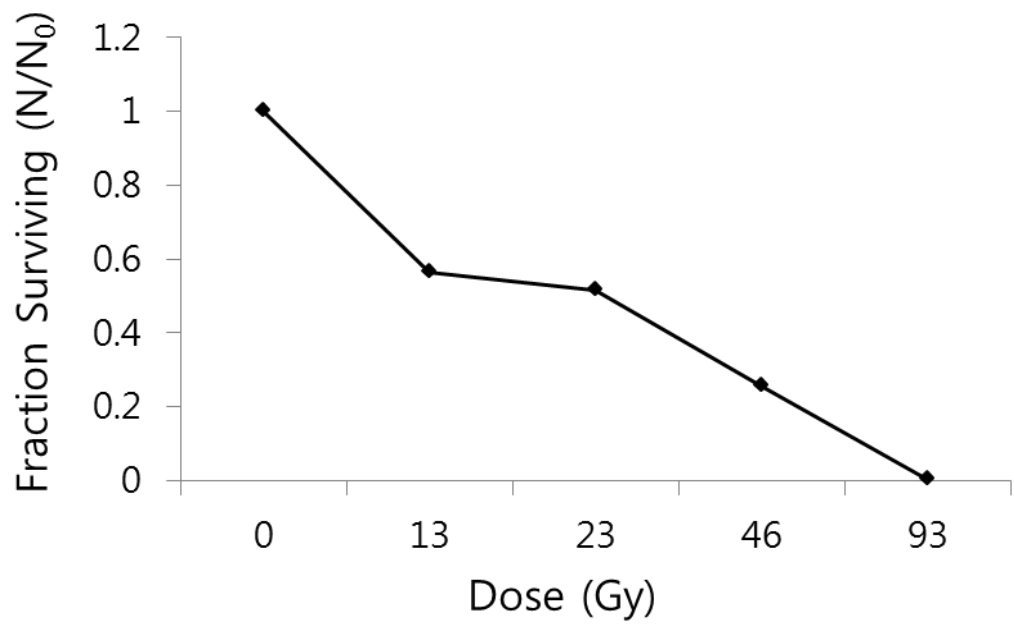

Figure 2. Survival fraction of E. coli irradiated with high-energy protons.

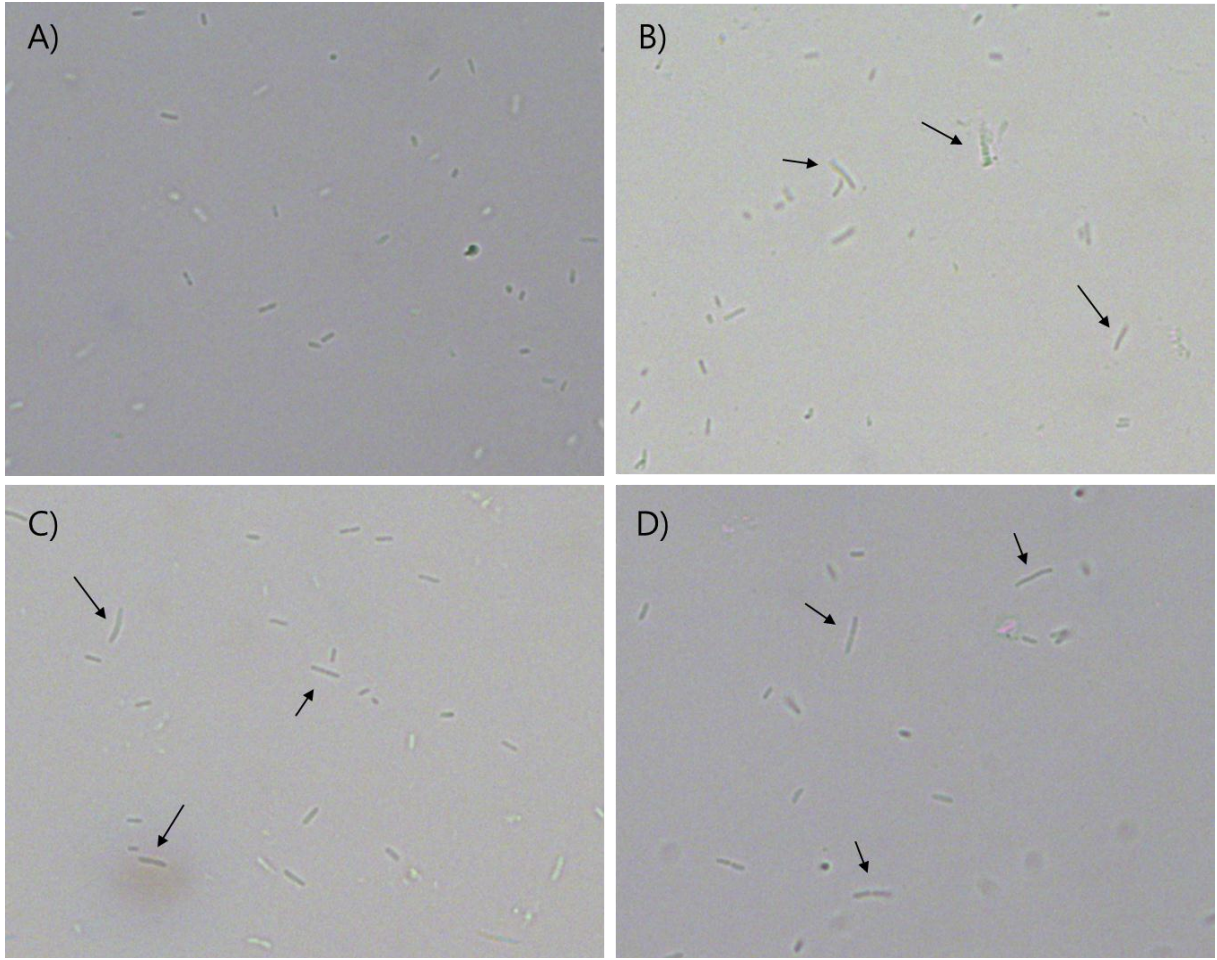

Figure 3. Optical microscopy of E. coli exposed to proton beams at dose ranging from 0 (A), 13 (B), 46 (C), and 93 Gy (D). Elongated E. coli was indicated with arrows (magnification of 400×).

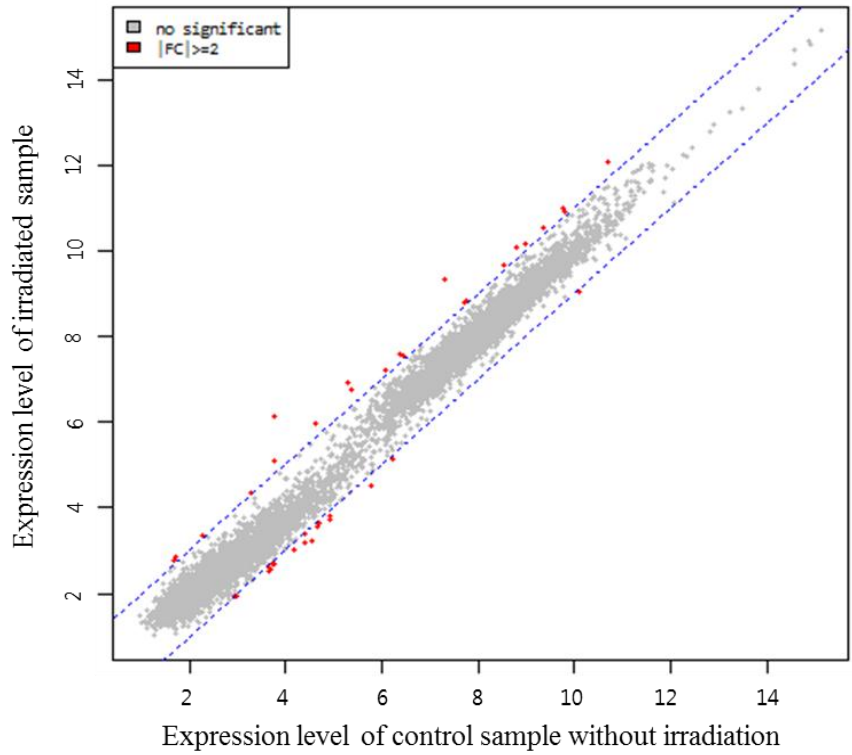

Figure 4. A scatter plot of microarray analysis with mRNA extracted from control samples and irradiated samples (93 Gy). Genes with no significant changes were indicated by gray spots and genes with greater than two fold changes were represented by red spots.

Table 1. The up-regulated genes (more than twofold) in irradiated samples (93 Gy) compared with control samples without irradiation

| Gene Symbol | Fold Change | Gene Title |
|---|---|---|
| c4973 | 4.059874 | hypothetical protein |
| aroG | 2.174009 | phospho-2-dehydro-3-deoxyheptonate aldolase |
| htpX | 2.132298 | heat shock protein HtpX |
| tnaA | 2.646636 | tryptophanase |
| phnH | 2.303034 | carbon-phosphorus lyase complex |
| rpmC | 2.175995 | 50S ribosomal protein L29 |
| tnaC | 5.106904 | tryptophanase |
| aceB | 2.146506 | malate synthase |
| rpsK | 2.318835 | 30S ribosomal protein S11 |
| c1304 | 2.119526 | hypothetical protein |
| ompW | 2.121414 | outer membrane protein W |
| yjcB | 2.206998 | hypothetical protein |
| gadY | 2.122184 | ncRNA |
| c4435 | 2.538039 | hypothetical protein |
| phnL | 2.654647 | predicted ATP transporter ATP-binding protein |
| phnJ | 2.108218 | phosphonate metabolism protein |
| glpK | 2.42282 | glycerol kinase |
| c3402 | 2.092045 | hypothetical protein |
| c4243 | 3.063662 | hypothetical protein |
| mutL | 2.508787 | methyl-directed mismatch repair protein |
| ompC | 2.312506 | outer membrane porin protein C |
| aceA | 2.279488 | isocitrate lyase |